\documentclass[12pt,fullpage]{article}

\newcommand{\be}{\begin{equation}}
\newcommand{\ee}{\end{equation}}
\newcommand{\beq}{\begin{eqnarray}}
\newcommand{\eeq}{\end{eqnarray}}
\newcommand{\AmS}{{\protect\the\textfont2

A\kern-.1667em\lower.5ex\hbox{M}\kern-.125emS}}

\advance\voffset by -1cm
\advance\hoffset by -.5cm
       
\begin{document}
\begin{flushright}
{TIFR/TH/05-16} \\
{July 19, 2005} \\
\end{flushright}

\vspace{1em}

\begin{center}
{\large{Hidden Variables, Non Contextuality and Einstein-Locality in 
Quantum Mechanics}} \\ [5mm]
{Virendra Singh} \\ 
{Tata Institute of Fundamental Research, Mumbai 400 005, India}
\end{center}

\vspace{2em}

\begin{enumerate}
\item[{1.}] The Formalism of Quantum Mechanics: \\
(a) States, (b) Physical Observables, (c) Dynamics, (d) Statistical Postulate.
\item[{2.}] Why Hidden Variables in Quantum Mechanics
\begin{enumerate}
\item[{2.1}] Determinism
\item[{2.2}] The Problem of Measurement and Incompleteness
\begin{enumerate}
\item[{2.2.1}] Bohr : Quantum System-Classical Apparatus
\item[{2.2.2}] Von-Neumann : Wave function Collapse
\item[{2.2.3}] Non Local Correlations
\end{enumerate}
\end{enumerate}
\item[{3.}] Von-Neumann's Proof of the Impossibility of Hidden Variable 
Theories
\begin{enumerate}
\item[{3.1}] Proof 
\item[{3.2}] Homogeneous Ensembles
\item[{3.3}] Reactions to Von-Neumann's Theorem
\end{enumerate}
\item[{4.}] Bells' Hidden Variable Model for a Spin One-half Particle
\begin{enumerate}
\item[{4.1}] The Model
\item[{4.2}] Analysis of Von-Neumann Theorem
\item[{4.3}] Jauch-Piron Impossibility Proof. 
\end{enumerate}
\item[{5.}] Non Contextuality and Quantum Mechanics
\begin{enumerate}
\item[{5.1}] Gleason's Theorem and Bell's Proof of it
\item[{5.2}] Non Contextuality 
\item[{5.3}] Kochen-Specker Theorem 
\item[{5.4}] Mermin's Examples 
\item[{5.5}] Some Other Aspects 
\begin{enumerate}
\item[{5.5.1}] Stochastic Noncontextual Hidden Variable Theories
\item[{5.5.2}] Finite Precision Measurements and Kochen-Specker Theorem
\end{enumerate}
\end{enumerate}
\item[{6.}] Einstein-Locality and Quantum Mechanics
\begin{enumerate}
\item[{6.1}] The Background
\item[{6.2}] Einstein-Podolsky-Rosen Theorem
\item[{6.3}] Einstein Locality
\item[{6.4}] Bohm's Version of the EPR Analysis
\item[{6.5}] Bell's Inequalities and Bell's Theorem
\item[{6.6}] Clauser-Home-Shimony-Hoft (CHSH) form of Bell's Inequalities
\item[{6.7}] Wigner's Proof of Bell-CHSH Inequalities
\item[{6.8}] Experimental Tests of Bell Inequalities
\item[{6.9}] Bell's Theorem without Inequalities
\begin{enumerate}
\item[{6.9.1}] Greenberger-Horne-Zeilinger (GHZ) Proof
\item[{6.9.2}] Hardy's Version of EPR Correlations
\end{enumerate}
\item[{6.10}] Superluminal Signalling
\item[{6.11}] More Bell's Inequalities
\end{enumerate}
\item[{7.}] Envoi
\item[{8.}] Bibiliographical Notes
\end{enumerate}

\newpage

\section{The Formalism of Quantum Mechanics}

There is general agreement on the basic formalism of Quantum Mechanics
which one uses in its physical applications. Briefly we can
recapitulate it as follows:

\noindent{\bf (a)~~The States :} 

The state of a quantum system is specified by a state vector
$|\psi>$. Further

(i) The state vectors $|\psi>$ and $|\psi'> = \exp (i\theta) |\psi>$,
where $\theta$ is a real constant, represent the same physical state,

(ii) Let $|\psi_1>$ and $|\psi_2>$ specify two different possible
states of a system $S$, then the linear combination,
\[
a |\psi_1> + b |\psi_2> ,
\]
where $a$ and $b$ are complex coefficients, also represents a possible
state of the system $S$.

(iii) The possible state vectors of a system $S$ belong to a Hilbert
space ${\cal H_S}$ equipped with a complex scalar product. The scalar
product of a state vector $|\psi>$ with the state vector $|\phi>$ is
denoted by $<\phi |\psi>$ and satisfies the relation
\[
<\phi |\psi>^\ast = <\psi |\phi> .
\]
The norm of a state $|\psi>$, $||\psi||$, is given by
\[
||\psi ||^2 = <\psi | \psi> .
\]
We shall normally use normalized $|\psi>$'s ie we take $||\psi||
=1$. 

\noindent{\bf (b)~~Physical Observables :} 

Any physical observable ${\cal A}$, of a system $S$, is represented by
a unique linear self-adjoint operator $A$ acting on $|\psi> \subset
D(A) \subset {\cal H}_S$, where $D(A)$ is called the domain of the
operator $A$.

The spectrum of the operator $A$, spec $(A)$, is the set of all those
complex numbers $z$ for which the operator $(A - zI)$ cannot be
inverted. For self-adjoint operators in Hilbert space it is a set of
real numbers $a_j\epsilon$ spec $(A)$. Depending on $A$, spec $(A)$
can be a discrete set, or a continuous set or a mixture of both. Any
$A$ measurement of a physical observable ${\cal A}$, always results in
one of real numbers $a_j$ occuring in spec $(A)$. Let
\[
A | a_j > = a_j | a_j > , \qquad\quad a_j \epsilon spec (A) .
\]
For the sake of brevity we will refers to $a_j$'s as eigenvalues of
$A$ and the $|a_j >$'s as the corresponding eigenvectors belonging to
the eigenvalue $a_j$ of $A$. All the eigenvectors of $A$ together form
a complete set and we can choose them to be an orthonormal set ie
\[
< a_j | a_k > = \delta_{(a_j, a_k)} 
\]
where $\delta_{(a_j, a_k)}$ is Kronecker delta symbol (for discrete
eigenvalues) or appropriate Dirac delta distribution (for continuous
part of the spectrum). The completeness relation can be stated as 
\[
\sum_{a_j} |a_j > < a_j | = {\bf 1}
\]
where the sigma symbol over $a_j$ stands for summation over discrete
part of the spectrum $a_j \epsilon$ spec $(A)$ and for an integral,
with proper measure, for the continuous part of the spectrum.

\noindent{\bf (c)~~Dynamics :} 

The dynamical evolution of the state vector $|\psi (t) >$, of a
systems, with time $t$, is given by the Schr\"odinger equation
\beq
i\hbar {\partial \over \partial t} |\psi (t) > &=& H \psi (t) \nonumber \\
{\rm ie} ~~~ \psi (t) &=& U (t - t') \psi (t') , \nonumber \\
{\rm with} ~~~ U(t) &=& \exp (-i Ht/\hbar) \nonumber 
\eeq
where $H$ is the Hamiltonian operator for the system $S$,
corresponding to it's physical energy observable. This evolution is
valid for a free evolution of the system between it's preparation and
subsequent measurement.

\noindent{\bf (d)~~Statistical Postulate :} 

If we measure a physical observable ${\cal A}$, on a system $S$, in
the state $|\psi>$, then the probability of measurement resulting in
the value $a_j \epsilon spec (A)$ is given
\[
prob ({\cal A}, a_j, |\psi>) = | <a_j |\psi> |^2 .
\]
Using this expression the expectation value $E_\psi ({\cal A})$, of
${\cal A}$ in the state $|\psi>$ is given by
\[
E_{|\psi>} ({\cal A}) = \sum_{a_j} a_j |<a_j |\psi>|^2 = <\psi |A|\psi> .
\]

If we define the projection operator $P(|\phi>)$ corresponding to the
state $|\phi>$ by
\[
P(|\phi>) = |\phi> <\phi| = P_\phi
\]
then we can write
\[
E_{|\psi>} ({\cal A}) = {\rm Tr} (A |\psi> <\psi|) = {\rm Tr} (A P_\psi), 
\]
and
\begin{eqnarray*}
{\rm prob} \ ({\cal A}, a_j, |\psi>)&=&{\rm Tr} (|a_j ><a_j |\psi><\psi|)\\
&=& {\rm Tr} (P_{a_j} P_\psi) . 
\end{eqnarray*}
We shall refer to the projection operator $|\psi><\psi|$ when the
system is in the state $|\psi>$ as it's density operator $\rho$ ie
\[
\rho = |\psi ><\psi| .
\]
If, however, the system is not in a pure state but we know that it is
in mixture of states $|\psi_k>$ with probability $p_k$, then it has
to be represented by the density matrix
\[
\rho = \sum_k p_k |\psi_k> <\psi_k |
\]
with $\sum_k p_k = 1, p_k \geq 0$. We now have 
\begin{eqnarray*}
{\rm prob} ({\cal A}, a_k; |\psi_k>, p_k)& =& {\rm Tr} (P_{a_k} \rho)\\
E_{[|\psi_k>, p_k]} ({\cal A}) &=& {\rm Tr} (A \rho) .  
\end{eqnarray*}

\section{Why Hidden Variables in Quantum Mechanics ?}

\subsection{Determinism} 

In the formalism of the quantum mechanics the state vector $|\psi>$
specifies the state of the system completely. If we now measure a
physical observable ${\cal A}$ for the system in the state $|\psi>$ we
obtain a value $a_k$, which is one of the possible eigenvalues of
$A$. The statistical postulate tells us about the probability with
which a particular value $a_k$, belonging to the spec $(A)$, will
occur. We are not able to predict as to which of the possible
eigenvalues $a_k$ of $A$ will occur. It is as if we had not specified
our state of the system completely by a knowledge of the state vector
$|\psi>$. The quantum mechanical state vector $|\psi>$ of a system
seems to encode only the statistical information.

Before the discovery of quantum mechanics, the statistical description
had been used in classical physics. For example, in classical
statistical mechanics we use such a description with great success,
to describe thermodynamic states of a classical system. These states
are a complicated average over microscopic states of the system which
are well defined in terms of positions and momentum coordinates of the
particles composing the system. In thermodynamics we are simply not
interested in such a detailed description of the system. Besides such a
detailed description, since it deals with extremely large number of
variables, is not always feasible.

Another example of random motion in classical physics is provided by
the Brownian motion of pollen grains in a fluid. We know here that the
randomness of the motion of pollen grains is due to their being
constantly buffeted by the large number of molecules of the fluid. The
molecules of the fluid remain hidden to our view under the microscope
used to observe pollen grain. If we include these molecules in the
dynamics then the motion is deterministic and not random. Incidently
the Einstein's expression for the root mean square displacement of a
Brown particle, divided by the time duration, has the same structure
as Heisenberg's uncertainty principle except that it involves
properties of fluid, such as diffusion coefficient, instead of
Planck's constant, in it.

In both these examples the systems for which we gave statistical
description in classical physics were not inherently statistical but
were deterministic. The statistical or random element arose because of
our averaging over a large number of variables which we ignored ie
which remained hidden. Could it be that the statistical description
given by quantum mechanics could also be arising from the same cause? 
We could then try to supplement the quantum mechanical specification
of the state of a system by it's state vector $|\psi>$ by adding extra
number of variables, referred to as ``hidden variables'', to provide a
more complete deterministic description of the system.

\subsection{The Problem of Measurement and Incompleteness} 

\subsubsection{Bohr : Quantum System-Classical Apparatus Split}

According to Niels Bohr the language of classical physics is the only
language available to us to describe experimental results. Classical
physics is thus needed to describe the results of experiments even on
the quantum systems which are usually microscopic. The quantum system
obeys laws of quantum physics. In contrast the measuring apparatus,
which are generally macroscopic, obey laws of classical physics. There
is thus a split of the world into system, which obeys quantum laws,
and apparatus, which obeys classical physics laws.
 
There is a problem of principle here. how much of the `system plus
apparatus' is to be put on quantum side ie into `system' and how much of
it is to put on classical side ie into `apparatus'. This is not a
severe problem in practice in as much as we put enough of `system +
apparatus' on the `system' side so that we achieve sufficient degree
of accuracy of description. The convergence is supposed to be ensured
by the folkloric ``correspondence principle'' according to which
quantum mechanics, in ``appropriate'' limits, goes into classical
physics. But in principle the split remains.

This split of the world into a microscopic quantum system and a
mesoscopic classical apparatus has been found deeply disturbing by
many persons. Is there a definite boundary between quantum and
classical at some mesoscopic level, which we will discover. This
possibility looks unlikely in view of the existence of macroscopic
quantum systems like squids. It is more likely that there is no
boundary at all. As John Bell says ``It is hard for me to envisage
intelligible discourse about a world with no classical part -- no base
of given events, be they only mental events in a single consciousness,
to be correlated. On the other hand, it is easy to imagine that the
classical domain could be extended to cover the whole. The wave
function would prove to be a provisional or incomplete description of
the quantum mechanical put, of which an objective account would be
come possible. It is this possibility of a homogeneous account of the
world, which is for me the chief motivation of the study of the so
called `hidden variable' possibility''.

\subsubsection{Von-Neumann : Wavefunction Collapse}

It would be more satisfactory to describe both the system and
apparatus by the same quantum laws especially if quantum mechanics has
pretensions to be the fundamental theory of physics. John Von-Neumann
regarded both the system and apparatus to be so describable by quantum
mechanics. He postulates a measurement interaction such that the
energy-coupling between the system, in a pure state $|\psi_n>$ of an
observable $\Omega$ (ie $\Omega |\psi_n> = \omega_n |\psi_n>$) with
eigenvalue $\omega_n$, and apparatus, for measuring the observable
$\Omega$, in a pure state $|f (a)>$ where $a$ is the pointer reading,
causes the system-apparatus to go from the initial state $|\psi_n>
|f(a)>$ to the final state $|\psi> |f (a_n)>$. The pointer-reading
$a_n$ thus indicates that the system was having the eigenvalue
$\omega_n$ of the observable $\Omega$. We shall further restrict
ourselves to measurement interaction of the type such that
\[
|\psi_n> |f(a)> \longrightarrow |\psi_n> |f(a_n) > .
\]

If the system is not initially in a pure state of the observable
$\Omega$ by is in a linear combination of such states ie the initial
state of the system $|\psi_{initial}>$ is given by
\[
|\psi_{initial}> = \sum_n c_n |\psi_n> ,
\]
the linearity of quantum mechanics leads to the system apparatus initial state 
\[
|\psi_{initial}> |f(a)> 
\]
to evolve into the final state 
\[
\sum_n c_n |\psi_n> |f (a_n)> .
\]
Thus the apparatus is in a superposition of states corresponding to
various pointer-readings and the probability of opening a pointer
reading $a_N$ given by $|c_N|^2$.

How does it happen that apparatus gives us, at the end of measurement,
a definite pointer reading say $a_N$ ie what causes the collapse of
the wave function
\[
\sum_n c_n |\psi_N> |f(a_N)> \ to \ |\psi_N> |f (a_N)> 
\]
with probability equal to $|c_N|^2$. According to Von-Neumann this is
caused by a casual and discontinuous process, called ``process of the
first kind'' by him, which take place only when a measurement is
completed. These are quite different by what he terms as ``processes
of the second kind'' which are causal and continuous and obey
Schr\"odinger equation, and by which a quantum system evolves in time
between its preparation and a measurement on it. Incidentally the
preparatus of a system in a definite state can also be regarded as a
special kind of measurement.

It is abundantly clear that even here quantum mechanics is not a
closed system since it has to be supplemented by ``processes of the
first kind'' which cause the collapse of a wavefunction at the end of
a measurement. It could be that we have to add extra dynamical hidden
variables to explicate these processes.
 
\subsubsection{Non-local Correlations}

The Einstein-Podolsky-Rozen (EPR) correlations are quite nonlocal. Let
two particles, each having spin ${1\over 2} \hbar$, in a simplest state
separate, one moving to the left and the second moving to the
right. Then a measurement of any component $\sigma_1 \cdot \hat{a}$ of
one of them allows us to know what $\sigma_2 \cdot \hat{a}$ would be if
measured even if these two particles are extremely for apart. It is
conceivable, in principle, that this nonlocal correlation comes about
through ``local causal'' correlations through the involvement of hidden
variables. If it turns out to be possible then one would not have to
invoke non-local causal mechanics. We will be discussing EPR
correlations in detail later on.  

\section{Von-Neumann's Proof of the Impossibility of Hidden-Variable Theories}

\subsection{Proof} John Von-Neumann published, in 1932, a result which
seemed to say that it is impossible to have a hidden variable
completion of the quantum mechanics. This result, given the immense
prestige of Von-Neumann as a mathematician, effectively blocked any
search for such theories for a long time. We therefore look at this
proof at this stage before proceeding any further.

The assumptions that go in the proof of Von-Neumann's theorem are as
follows. We use Gothic letters to denote physical observables and the
corresponding Latin symbol for the Hermitian operator associated with
it in quantum mechanics e.g. for the physical observable $\Re$ the
Hermitian operator associated is denoted by $R$. The expectation value
of $\Re$ is denoted by $<\Re>$ in the physical ensemble.
\begin{enumerate} 
\item[{(0)}] To every physical observable ${\cal A}$
there corresponds only \underbar{one} Hermitian operator $A$. We use
the notation ${\cal A} \mapsto A$ for this correspondence. 

\item[{(I)}] If a physical observable $\Re$ is non-negative then it's 
expectation value $<\Re>$ is also non-negative.  

\item[{(II)}] If $\Re_1 + \Re_2$ are two physical observables then the 
physical observable associated with their sum $\Re_1 + \Re_2$ is associated
with the Hermitian operator $R_1 + R_2$.

\item[{(III)}] If $\Re_1, \Re_2, \cdots $ are physical observables and 
$a_1, a_2, \cdots$ real numbers then \[ \langle (a_1 \Re_1 + a_2 \Re_2
+ \cdots) \rangle = a_1 <\Re_1> + a_2 <\Re_2> + \cdots .  \]

\item[{(IV)}] If a physical observable $\Re$ is associated with a
Hemitian operator $R$ then a function $f(\Re)$ is associated with the
Hermitian operator $f(R)$.  
\end{enumerate}

Von-Neumann first proves, using only assumptions (0)--(III), the
following theorem.

\noindent\underbar{Theorem}: There exists a density operator $\rho$ such that 
\[
<\Re> = {\rm Tr} (\rho R)
\]
for all $\Re$. here the density operator depends only on the ensemble
of physical states but is independent of the physical observable
$\Re$. Further
\[
{\rm Tr} \rho = 1 .
\]
\underbar{Proof}: Given a complete orthonormal set $|n>$, ie
\[
<n|m> = \delta (n,m) \ \ \ and \ \ \ \sum_n |n> <n| = 1.
\]
We can express any hermitian operator $R$, as 
\[
R = \sum_{n,m} |n> <m| R_{nm}
\]
where $R_{nm} = <n |R| m>$. Further we have
\[
R_{mn}^\ast = R_{nm} = <m |R| n>^\ast .
\]
We can rearrange various terms to rewrite
\[
R = \sum_n U_{nn} R_{nn} + \sum_{n>m} V_{nm} Re R_{nm} + \sum_{n>m}
W_{nm} {\rm Im} R_{nm} 
\]
where
\beq
U_{nn} &=& |n> <n| \nonumber \\
V_{nm} &=& |n> <m| + |m> <n| \nonumber \\
W_{nm} &=& i [|n> <m| - |m> <n|] \nonumber 
\eeq
The operators $U_{nn}, V_{nm}$ and $W_{nm}$ are all hermitian and will
be taken to represent the observables according to the correspondence,
(assumption (0)), 
\beq
{\cal U}_{nn} &\longmapsto& U_{nn} \nonumber \\
\Upsilon_{nm} &\longmapsto& V_{nm} \nonumber \\
{\cal W}_{nm} &\longmapsto& W_{nm} . \nonumber 
\eeq
By assumption (0) and (II), we have an observable $\Re$ given by
\[
\Re = \sum_n {\cal U}_{nn} R_{nn} + \sum_{n>m} \Upsilon_{nm} Re R_{mn}
+ \sum_{n>m} {\cal W}_{nm} {\rm Im} R_{nm} 
\]
We thus have the correspondence
\[
\Re \longmapsto R .
\]
By assumption III, the expectation value $<\Re>$ of $\Re$ in the
ensemble corresponding to a quantum state is given by
\[
<\Re> = \sum_n <{\cal U}_{nn}> R_{nn} + \sum_{n>m} <\Upsilon_{nm}> Re R_{nm}
+ \sum_{n>m} <{\cal W}_{nm}> {\rm Im} R_{nm} .
\]
We now introduce the notation, 
\beq
\rho_{nn} &=& <{\cal U}_{nn}> , \nonumber \\
\rho_{nm} &=& {1\over 2} \left[ <\Upsilon_{nm}> + i <{\cal W}_{nm}> \right] \ \
{\rm for} \ n > m , \nonumber \\
\rho_{mn} &=& {1\over 2} \left[ <\Upsilon_{nm}> - i <{\cal
W}_{nm}>\right]  \ \ {\rm for} \ n > m . \nonumber 
\eeq
In terms of the expressions $\rho$, we can express
\[
<\Re> = \sum_{n,m} \rho_{mn} R_{nm} .
\]

We now define a hermitian density operator $\rho$ by
\[
<m | \rho | n > = \rho_{mn}
\]
Using the density operator $\rho$, we can finally express
\[
<\Re> = {\rm Tr} (\rho R) .
\]
Note that $\rho$ is independent of $\Re$ but only depends on the
ensemble. 

For the constant unit observable $\Re = 1 \mapsto R = 1$, we get
\[
{\rm Tr} \rho = 1 .
\]

Von-Neumann next uses projection operators to show that the density
operator is positive semidefinite. Consider the correspondence 
\[
{\cal P}_\phi \longmapsto |\phi > < \phi |
\]
By (IV), we get
\[
({\cal P}_\phi)^2 \longmapsto (|\phi> <\phi|)^2 = |\phi> <\phi| 
\]
we thus get
\beq
<{\cal P}^2_\phi> &=& <{\cal P}_\phi> = {\rm Tr} (\rho {\cal P}_\phi) =
{\rm Tr} (\rho |\phi> <\phi|) \nonumber \\
&=& <\phi |\rho| \phi> \nonumber  
\eeq
By (1), it follows that 
\[
<{\cal P}^2_\phi> \geq 0 
\]
and therefore
\[
<\phi |\rho| \phi> \geq 0 .
\]
\underbar{Theorm}: The density operator $\rho$ is hermitian and
positive semidefinite. 

We now proceed to prove the Von-Neumann theorem on the absence of
hidden variable in quantum mechanics using these results and
assumption (IV). We are going to show that the dispersion-free states
do not exists. If they do than for them, we must have 
\[
< \Re^2 > = < \Re >^2 .
\]
By assumption (IV) if
\[
\Re \longmapsto R 
\]
then
\[
\Re^2 \longmapsto R^2 .
\]
The dispersion free condition then becomes
\[
{\rm Tr} (\rho R^2) = [{\rm Tr} (\rho R)]^2. 
\]
In Particular for $R = |\phi> <\phi| = P_\phi$ 
\[
{\rm Tr} (\rho P^2_\phi) = [{\rm Tr} (\rho P_\phi)]^2 
\]
using $P^2_\phi = P_\phi$ we get
\[
{\rm Tr} (\rho P_\phi) [ 1 - {\rm Tr} (\rho P_\phi)] = 0 
\]
ie
\[
{\rm Tr} (\rho P_\phi) = <\phi |\rho| \phi>
\]
can be either 0 or 1. If we take
\[
\phi = \cos \theta \phi_1 + \sin \theta \phi_2 
\]
then $<\phi |\rho| \phi>$ varies continuously with $\theta$. Thus we
have only the following two possibilities left
(i) $<\phi |\rho| \phi> = 0$ for all $\phi$ or (ii) $<\phi |\rho| \phi>
= 1$ for all $\phi$. Thus either $\rho = 0$ or $\rho = 1$. Both these
possibilities are unacceptable physically. Hence no dispersion-free
states can exist. 

As John Von-Neumann concluded ``It is therefore not, as is often
assumed a question of reinterpretation of quantum-mechanics, the
present system of quantum-mechanics would have to be objectively
false, in order that another description of the elementary process
than the statistical one be possible''. 

\subsection{Homogeneous Ensembles} 

An ensemble characterized by a density matrix $\rho$ is called
homogeneous, if whenever we write 
\[
\rho = \rho_1 + \rho_2 
\]
and where 
\[
\rho_1 = \rho_1^\dagger \ , \ \ \rho_1 \geq 0 \ , 
\]
and
\[
\rho_2 = \rho_2^\dagger \ , \ \ \rho_2 \geq 0 
\]
then
\[
\rho_1 = c_1 \rho \ , \ \ \rho_2 = c_2 \rho 
\]
where
\[
c_1 \geq 0 \ , \ \ c_2 \geq 0 \ , \ \ c_1 + c_2 = 1 .
\]
Von-Neumann showed that $\rho$ corresponding to a homogeneous ensemble
has to be some projection operator ie $\rho = P_\phi$ for some
$\phi$. We thus have for homogeneous ensembles
\[
<\Re> = {\rm Tr} (R P) = <\phi |R| \phi> 
\]
which is the usual quantum-mechanical expectation value. 

\subsection{Reactions to Von-Neumann's Theorem}

After it's publication in 1932, Von-Neumann's theorem soon acquired
the status of a dogmatic pronouncement against hidden variable
theories. At a conference on ``New Theories of Physics'' at Warsaw in
1938, Niels Bohr, the guiding sprit behind the long-dominant
Copenhagen interpretation of Quantum mechanics, publicly endorsed it
after a presentation of it by Von-Neumann. Bohr expressed his
admiration for the result. He also mentioned that the conclusion one
of his own papers was essentially the same even though he arrived at
it by using more elementary ways. 

Max Born, in his book ``Natural Philosophy of Cause and Chance'',
first published in 1949, said ``The result is that the formalism of
quantum mechanics is uniquely determined by these axions; in particular
no concealed parameters can be introduced with the help of which the
indeterministic description could be transformed into a deterministic
one. Hence if a future theory should be deterministic, it can not be a
modification of the present one but must be essentially different. How
this should be possible without sacrificing a whole treasure of
well-established results I leave to the determinist to worry
about''. This quotation is from an English translation of Born's 1949
book which was published as late as in 1964. 

Even more telling are the recollection of Paul Feyeraband, the well
known philosopher of science, of a public lecture and seminar by Bohr
in Askov (sometime during 1949-1952). He says ``In Askov, I also met
Niels Bohr. He came for a public lecture and conducted a seminar, both
in Danish ...... At the end of the lecture he left, and the discussion
proceeded without him. Some speakers attacked his qualitative
arguments --- there seemed to be lots of loopholes. The Bohrians did
not clarify the arguments; they mentioned an alleged proof by Von
Neumann, and that settled the matter. Now I very much doubt that those
who mentioned the proof, with the possible exception of one or two of
them, could have explained it. I am also sure their opponents had no
idea of it's details. yet, like magic, the mere name ``Von Neumann'
and the me word ``proof'' silenced the objectors. I found this very
strange but was relieved to remember that Bohr himself had never used
such tricks''. 

Despite the fact that in 1952 David Bohm actually published a hidden
variable model of quantum mechanics, the spell of von-Neumann's
theorem remained unbroken. The ``hidden-variables'' in Bohm's model
were particle trajectories. It was a resurrection of earlier
de-Broglie pilot wave theory model of 1927 which had been abondoned by
its' originator under criticism by Pauli and Einstein. Bohm was
however able to take care of various criticisms of Pauli, Einstein and
of de Broglie, besides supplementing it with a theory of
measurements. Bohm however did not analyze Von-Neumann's proof to
pinpoint as to how he was able to circumvent Von-Neumann's theorem. He
must have clearly violated some assumption involved in it. Such as
analysis was first carried out by John Bell, in the context of his
hidden variable model for a spin one-half particle, in 1966. The spell
of Von-Neumann's theorem was at last broken as a result of Bell's
work. 

\section{Bell's Hidden Variable Model for a Spin One-half Particle} 

\subsection{The Model}

We consider, following Bell, a quantum mechanical spin one-half
particle and ignore its translational degrees of freedom. It's
description requires a two dimensional Hilbert space. The most general
observable $M$ is specified by four real numbers, $\alpha$ and
$\vec\beta$, and can be taken to be
\[
M (\alpha, \vec\beta) = \alpha \underline{1} + \vec\beta \cdot
\vec\sigma 
\]
where $\sigma$'s are usual three Pauli-matrices and 1 is a unit
matrix. The eigenvalues of $M(\alpha, \beta)$ are given by $(\alpha +
|\vec\beta|)$ and $(\alpha - |\vec\beta|)$. The physical observable
$M$ can take only these two eigenvalue as it's values. 

The state of this system is in quantum-mechanics described by a state
vector $|\psi>$ which we take to be normalised ie $<\psi | \psi> =
1$. We now add a hidden variable $\lambda$, which is a real variable
having the range ${1\over 2} \geq \lambda \geq - {1\over 2}$, to the
quantum state vector $|\psi>$ to complete the specification of
dispersion free states. 

For the dispersion free state $(|\psi>, \lambda)$ we associate the
value for the observable $M(\alpha, \vec\beta)$ given by 
\[
M(\alpha, \vec\beta, |\psi>, \lambda) = \alpha + |\vec\beta| sgn
[<\psi| \vec\beta \cdot \vec\sigma |\psi>] \times sgn \{\lambda
|\vec\beta| + {1\over 2} |<\psi |\vec\beta \cdot \vec\sigma|
\psi>|\}. 
\]
Clearly $M(\alpha, \vec\beta, |\psi>, \lambda)$ only taken the values
$\alpha \pm |\vec\beta|$ in the dispersion free states $(|\psi>,
\lambda)$ as is required. 

The ensemble of dispersion free states is taken be one in which the
hidden value $\lambda$ is distributed with equal probability over it's
range ${1\over 2} \geq \lambda \geq - {1\over 2}$. We shall now check
whether the ensemble average $<M>$ over this ensemble reproduces the
quantum mechanical expectation value of $M(\alpha, \vec\beta)$ which
is given by
\[
<\psi | M (\alpha, \vec\beta, |\psi> .
\]
We therefore integrate $M (\alpha, \vec\beta, |\psi>, \lambda)$ over
$\lambda$ ie 
\[
<M> = \int^{1/2}_{-1/2} d\lambda \ M (\alpha, \vec\beta, |\psi>,
\lambda) .
\]
An easy calculation leads to 
\beq
<M> &=& \alpha + <\psi |\vec\beta \cdot \vec\sigma |\psi> \nonumber \\
&=& <\psi | M (\alpha, \vec\beta) |\psi> . \nonumber
\eeq
It is thus equal to the quantum mechanical expectation value of
$M(\alpha, \vec\beta)$. Bell thus succeeded in providing a hidden
variable model of a quantum translationless spin ${1\over 2}$
particle. 

\subsection{Analysis of Von-Neumann Theorem}

Now that we have an explicit example of the hidden variable model we
can use it to analyse the Von-Neumann's impossibility proof. The
guilty assumption turns out to the innocuous looking one III
\[
a_1 <{\cal A}_1> + a_2 <{\cal A}_2> = <a_1 {A}_1 + a_2 {A}_2> , 
\]
even when $A_1$ and $A_2$ are noncommuting Hermian operators. While it
is true of quantum mechanical states but is not true for dispersion
free states. After all the eigenvalues $\pm \sqrt{2}$ of the operator
$\sigma_x + \sigma_y$ are not sum of the eigenvalues $\pm 1$ of
$\sigma_2$ and eigenvalues $\pm 1$ of $\sigma_y$. A physical
explanation of this phenomenon is that we require three different
orientation of Stern-Gerlach magnets to measure $\sigma_x, \sigma_y$
and $(\sigma_x + \sigma_y)$. 

\subsection{Jauch-Piron Impossibility Proof}

Bell also analysed a new proof of impossibility of hidden variables in
quantum mechanics which was given by Jauch and Piron in 1963. They
deal with expectation values of only projection operators. For a
projection operator $P$, since $P^2 = P$ an eigenvalue can be only 0
or 1. The expectation value of $<P>$ thus gives us the probability
that we observe the eigenvalue 1 for it. 

Jauch and Piron assume
\[
<A> + <B> = <(A + B)> 
\]
only for commuting projection operators $A$ and $B$. To this extent it
is an improvement of Von-Neumann's work where this assumption was made
for all operators $A$ and $B$ whether they were commuting or
noncommuting. Given two projection operators $A$ and $B$ they also
define another projection operator $(A \cap B)$ which projects out the
intersection of the subspaces of $A$ and $B$. They further assume that
if, in some state, 
\[
<A> = <B> = 1 
\]
then $<A \cap B> = 1$ for that state. This assumption is by analogy to
logie of propositions in where the value 1 corresponds to `truth' and
value 0 to `falsehood'. It corresponds to logical proposition that if
$A$ is `true' and $B$ is `true' then ($A$ `and' $B$) is also `true'. 

Consider a two dimension subspace and the projection operators
${1\over 2} (1 + \vec a \cdot \vec\sigma)$ and ${1\over 2} (1 -
\vec a \cdot \vec\sigma)$ where $\vec a$ is a unit vectors. Since they
commute and add upto 1, we have 
\[
<{1\over 2} (1 + \vec\alpha \cdot \vec\sigma)> + <{1\over 2} (1 -
\vec\alpha \cdot \vec\sigma)> = <1> = 1 .
\]
Thus, for dispersion free states, either $<{1\over 2} (1 + \vec \alpha
\cdot \vec\sigma> = 1$ or $<{1\over 2} (1 - \vec \alpha \cdot \vec\sigma)>
= 1$. Similarly either 
\[
< {1\over 2} (1 \pm \vec\beta \cdot \vec\sigma)> = 1 \ \ {\rm or} \ \
<{1\over 2} (1 - \vec\beta \cdot \vec\sigma)> = 1
\]
for another unit vector $\vec\beta$. Thus by choosing $A$ to be either
${1\over 2} (1 + \vec\alpha \cdot \vec\sigma)$ or ${1\over 2} (1 -
\vec\alpha \cdot \vec\sigma)$, and $B$ to be either ${1\over 2} 
(1 + \vec\beta \cdot \vec\sigma)$ or ${1\over 2} (1 -
\vec\beta \cdot \vec\sigma)$ appropriately we can arrange
\[
<A> = <B> = 1
\]
But now $A \cap B = 0$, so that $<A \cap B> = 0$ which contradicts the
assumption made earlier. Hence dispersion free states cannot exist. 

The objection to Jauch-Piron proof, as Bell pointed out, is again that
we are dealing with measurements and not with logical
propositions. Their second assumption, while obeyed by quantum
mechanical states is not necessarily obeyed by the hidden variable
states. 

\section{Non Contextuality and Quantum Mechanics}

\subsection{Gleason's Theorem and Bell's Proof of it}

We saw that a major ingredient of Von-Neumann's impossibility proof of
hidden variables was his proof of existence of a Hermitian positive
semidefinite density matrix. Von-Neumann had to assume that
expectation value of a sum of given observables is a sum of
expectation values of these observable even if they are a noncommuting
set. Gleason (1957) was able to prove the existence of such a density
matrix by making the assumption of additivities of expectation values
only for commuting observables provided the dimension of the Hilbert
space for the system is greater than or equal to three. This would
allow us again to prove the impossibility of hidden variables in
quantum mechanics for systems with dim ${\cal H}_s \geq 3$. Bell's
counter exempts of spin one-half particle, since it's Hilbert space is
two dimensional, has no relevance for this case. 

We shall now proceed to reproduce Bell's proof (1966) of Gleason
result. Let a set $(\phi_1, \phi_2, \phi_3, \cdots )$ be orthogonal
and complete. The projection operators 
\[
P(\phi_i) = |\phi_i > < \phi_i | / <\phi_i | \phi_i > 
\]
commute with other ie
\[
[P (\phi_i), P (\phi_i)] = 0 .
\]
Further 
\[
\sum_i P (\phi_i) = 1 .
\]
Using the additivity assumption for the expectation value of commuting
observables only we get 
\be
\langle \sum_i P(\phi_i) \rangle = \sum_i \langle P (\phi_i) \rangle =
1 .
\ee
We also note, since any eigenvalue of a projection operator is either
0 or 1, 
\be
\langle P (\phi_i) \rangle \geq 0 .
\ee
We thus have two corollaries (A) and (B); \\
(A) If in a given state, we have $<P(\phi)> = 1$ then $<P(\phi')> = 0$
for $\phi'$ orthogonal to $\phi$ ie $<\phi' |\phi> = 0$. 

This corollary is obvious since we can always have a complete
orthonormal set containing $|\phi>$, $|\phi'>$ and using (1) and (2). 

(B) If, in a given state, we have 
\[
\langle P(\phi_1) \rangle = \langle P (\phi_2) \rangle =
0 \ \ {\rm for~some} \ \ \langle \phi_1 | \phi_2 \rangle = 0  
\]
then $\langle P(\alpha \phi_1 + \beta \phi_2) \rangle = 0$ 
for all $\alpha, \beta$. 

We note here 
\beq
0 &=& \langle P (\phi_1) \rangle + P (\phi_2) = 1 - \sum_{i \neq 1,2}
\langle P (\phi_i)\rangle \nonumber \\
&=& \langle P(\psi_1) \rangle + \langle P (\psi_2) \rangle \nonumber 
\eeq
where
\beq
\psi_1 &=& \alpha \phi_1 + \beta \phi_2 \nonumber \\
\psi_2 &=& - \beta \phi_1 + \alpha \phi_2 \nonumber 
\eeq
since $(\psi_1/||\psi_1||, \psi_2/||\psi_2||, \phi_3, \phi_4, \cdots)$
is also a complete set if $(\phi_1, \phi_2, \phi_3, \phi_4, \cdots)$
is the corollary (B) follows from the positivity of  $\langle
P(\psi_1)\rangle$ and $\langle P(\psi_2)\rangle$. 

The strategy now would be to first prove that if $<P(\psi)> = 1$ and
$<P(\phi)> = 0$ then $|\psi>$ and $|\phi>$ can not arbitrarily close
to each other or more precisely
\be
|| (\psi - \phi) || \geq k || \psi || \ \ \ {\rm with} \ \ k > 0 .
\ee
Note that the constant $k$ is strictly positive in (3). In fact we
will show that $k$ can be chosen to be $\sqrt{(2 - 4/\sqrt{5})}$. This
result will then be used to prove the absence of dispersion free
states. To prove the result in eqn. (3) we shall use repeatedly
corollaries (A) and (B). We now give the proof. Let
\[
\phi = \sqrt{1 - \epsilon^2} \psi + \epsilon \psi'
\]
where $<\psi' |\psi> = 0$. Let $\psi''$ be a state vector such that 
\[
\langle \psi'' |\psi \rangle = \langle \psi'' |\psi' \rangle = 0 , \ \
\langle \psi'' |\psi'' \rangle = 1 .
\]
It follow $<\phi |\psi''> = 0$. 

Using (A) we get 
\beq
&& \langle P (\psi'') \rangle = 0 \ \ \ {\rm since} \ \ \langle \psi''
|\psi \rangle = 0 , \nonumber \\
{\rm and} && \langle P (\psi') \rangle = 0 \ \ \ {\rm since} \ \
\langle \psi' |\psi \rangle = 0 . \nonumber 
\eeq

Now, using (B) we have
\[
\langle P (\phi + {\epsilon \over \gamma} \psi'' \rangle = 0 
\]
since $<P (\phi)> = <P(\psi'')> = 0$. Here $\gamma$ is a constant to be
chosen later. We also have, using (B), 
\beq
&& \langle P (- \epsilon \psi' + \gamma \epsilon \psi'') \rangle = 0
\nonumber \\
{\rm since} && \langle P (\psi') \rangle = \langle P (\psi'') \rangle
= 0 \nonumber 
\eeq

We now note that 
\beq
&& \langle - \epsilon \psi' + \gamma \epsilon \psi'' | \phi +
{\epsilon \over \gamma} \psi'' \rangle \nonumber \\
&& = - \epsilon \langle \psi' |\phi \rangle + \epsilon^2 = 0 \nonumber
\eeq
We can, using (B), see 
\beq
&& \langle P (\phi + {\epsilon \over \gamma} \psi'' \rangle + \langle
P (- \epsilon \psi' + \gamma \epsilon \psi'') \rangle \nonumber \\
&& = \langle P (\sqrt{1 - \epsilon^2} \psi + \epsilon (\gamma +
{1\over \gamma}) \psi'' ) \rangle = 0 . \nonumber 
\eeq
We can now choose, if $\epsilon^2 \leq {1\over 5}$, $\gamma$ such that
\[
(\gamma + {1\over \gamma}) \epsilon = \pm \sqrt{1 - \epsilon^2} .
\]
We thus get 
\[
\langle P (\psi \pm \psi'') \rangle = 0 .
\]
Using (B) once again we get
\[
\langle P (\psi) \rangle = 0 .
\]
Since $<P (\psi)> = 1$ we have a contradition. We must therefore have
$\epsilon^2 \geq {1\over 5}$. This leads to 
\[
|\psi - \phi | = \sqrt{2 -2 \sqrt{(1 - \epsilon^2)}} |\psi| \geq (2 -
{4\over \sqrt{5}})^{1/2} |\psi> .
\]
The proof needs at least three linearly independant vectors $\psi,
\psi', \psi''$ and thus the Hilbert space must dimension $\geq 3$. 

If there are dispersion freestates then each projector has expectation
value either 0 or 1. Consider the expectation value projector 
\[
P(\theta) = \langle P (\cos\theta \psi + \sin\theta \phi) \rangle . 
\]
We have $P(\theta) = 0$ or 1, $P(0) = 1, P({\pi \over 2}) = 0$. There
must thus be at least one $\theta$ value, say $\theta_0$, at which
$P(\theta)$ jumps from the value 1 to value 0. 

We thus have two states which are arbitrarily close such that for one
of them projector has expectation value equal to 1 and for the other
one it is zero. This contradicts the result derived above. We thus
have proved the impossibility of hidden variable theories assuming
only the innocuous looking assumption about additivity of expectation
values of only commuting physical observables. 

How did this happen? To trace the culprit, we note that we can have a
set of orthonormal states
\[
\{\phi_1, \phi_2, \phi_3, \cdots \}
\]
and another set of orthonormal states, also containing $\phi_1$, 
\[
\{\phi_1, \phi'_2, \phi'_3, \cdots \}.
\]
We would then have
\[
1 = P (\phi_1) + P (\phi_2) + P (\phi_3) + \cdots 
\]
with $[P (\phi_j), P(\phi_k)] = 0$ for $j \geq 1, k \geq 1$, and  
\[
1 = P(\phi_1) + P(\phi'_2) + P(\phi'_3) + \cdots 
\]
with 
\begin{eqnarray*}
[P(\phi_1), P(\phi'_k)] &=& 0 \ \ {\rm for} \ \ k \geq 2 ; \\
{\rm and} \ \ \ [P(\phi'_j), P(\phi'_k)] &=& 0 \ \ {\rm for} \ \ j \geq
2, k \geq 2.
\end{eqnarray*}
What is assumed implicitly in Gleason's proof is that measuring
$P(\phi_1)$ is independent of whether one is measuring $P(\phi_2),
P(\phi_3), \cdots$ along with it or one is measuring $P(\phi'_2),
P(\phi'_3), \cdots$ along with it. However in general, since
\[
[P(\phi_j), P(\phi'_k)] \neq 0 \ \ {\rm for} \ \ j \geq 2, \ k \geq 2.
\]
We cannot measure $P(\phi_j)$ and $P(\phi'_k) (j \geq 2, k \geq 2)$
together. We thus can raise physical objections against the Gleason's
proof as against Von-Neumann's proof. 

\subsection{Non Contextuality}

We now summarise the moral learned from Gleason-Bell proof. Let a
physical observable $A$ commute with $B_1, B_2 \cdots$ which mutually
commute with each other i.e.
\[
[A, B_j] = 0, \ [B_i, B_j] = 0 .
\]
We can measure $A$ along with $B_j$'s together. The set $[A, B_1, B_2,
\cdots]$ will be said to form a context. let there be another set $(A,
C_1, C_2, \cdots)$ of mutually commuting observables and it will form
another context. In general however $B_j$'s will not commute with
$C_k$'s ie
\[
[B_j, C_k] \neq 0
\]
and are thus not simultaneously observable. The measurement of $A$
thus should be expected to depend on the context. The culprit in
Bell-Gleason proof of nonexistence of hidden variable theories was
implicit assumption of \underbar{Non-contextuality}. As Bell said ``It
was tacitly assumed that measurement of an observable must yield the
same value independently of what other measurements may be made
simultaneously''. 

\subsection{Kochen-Specker Theorem}

Bell's proof involves a consideration of a continuum of projection
operators to prove ``non existence'' of hidden variable theories using
noncontextuality and additivity of expectation values of only
commuting projection operators. Philosophers of science refer to it as
a ``continuum proof''. They prefer another version which uses onlya
finite number of projection operators which was given by S. Kochen and
E. Specker in 1967. This proof is given for a spin-one quantum
particle which needs a Hilbert space of three dimensions. As one can
always embed a 3-dimensional Hilbert space in a higher dimensional
one, the result obtained remains valid for any system with Hilbert
space of three or more dimensions. 

Consider the angular-momentum vector $\vec J = (J_x, J_y, J_z)$ for a
spin-one particle. Using the standard commutation rules for angular
momentum operators, it is easy to see that the three
operators. $J^2_x, J^2_y, J^2_z$ commute with other and further
\[
J^2_x + J^2_y + J^2_z = \hbar^2 .
\]
We will set $\hbar = 1$ from now on. The eigenvalues of $J^2_x, J^2_y,
J^2_z$ are 0 and 1. It follows that, in a dispersion free state, out
of these three operators two of them must take the value 1 and the
third one the value zero. Otherwise they cannot sum up to two. We can
also recast this example in terms of three commuting projection
operators $P_x, P_y, P_z$ which sum to unity, and whose eigenvalues
are 0 and 1. Then, in a dispersion free state, one of the three
projection operator would take the value 1, while the other two would
take the value 0. As it looks more physical we will continue to take
in terms of squares of the components of the angular
momentum. Similarly if $(n_1, n_2, n_3)$ is triad of unit vectors
which are perpendicular to each other, than again $J^2_{n_1},
J^2_{n_2}, J^2_{n_3}$ commute with other and add upto 2. Further, in a
dispersion free state two of them must take the value 1. while the
third one should be zero. 

More playfully the problem is same as the following Kochen-Specker
colouring problem. Take a number of \underbar{orthogonal} triads and
start coloring them green if $S^2_n = 0$ or red if $S^2_n = 1$. In ech
triad $(n_1, n_2, n_3)$ two of vectors will be coloured red and the
third one green. Is it always possible to carry out this colouring job
consistantly without ever coming to impasse that the same vector has
to coloured both red and greed? Kochen and specker considered a set of
117 directions and showed that such is not the case. Their
construction is highly intricate geometrically. A Peres found, in
1991, a choice of particularly elegant choice of 33 directions which
also allow one to prove the same result. These are all the rays
obtained by taking the \underbar{squares} of three direction cosines
to be $(0, 0, 1), (0, {1\over 2}, {1\over 2}), (0, {1\over 3}, {2\over
3})$ and $({1\over 4}, {1\over 4}, {1\over 2})$. A ray with direction
cosines $(a, b, c)$ is regarded equivalent to one with $(-a, -b,
-c)$. A figure containing exactly these 33 directions occurs in one of
Escher's drawing of an impossible waterfalls on the top of one of the
towers. The record of the minimum directions needed for proof is 31
and is due to Conway and Kochen.

We thus see that it not possible for a non contextuatic hidden
variable theory of reproduce quantum mechanical results. 

\subsection{Mermin's Examples} 

The proofs of Kochen-Specker theorem involve more intricate geometry
than is to the taste of most physicists. Mermin considered a set of
nine physical observables for a system of two spin one-half
particles. For this system, with a Hilbert space having dimension
equal to 4, it is possible to prove Bell-Kochen-Specker result in a
very simple algebraic fashion using only the properties of
Pauli-matrices. 

Let $A, B, C, \cdots$ be a commuting set of observables. Let
eigenvalues of $A, B, C, \cdots$ be denoted by $a, b, c, \cdots$. Let
there exist a functional relationship between these commuting
observables is given by 
\[
f (A, B, C, \cdots ) = 0 .
\]
Let $v(A)$ be the value the observable $A$ in the dispersion free
state. We shall take $v (A)$ to be one of the eigenvalues $a$ of
$A$. Similarly for $v(B)$ etc. These values $v(A), v(B), \cdots$
must also satisfy
\[
f(v(A), v(B), v(C), \cdots) = 0 .
\]

The set of nine observables are arranged in a $3 \times 3$ array given
below: 
\[
\begin{matrix}
{ & V1 & V2 & V3 \cr &&& \cr H1: & \sigma^{(1)}_x & \sigma^{(2)}_x &
\sigma^{(1)}_x \sigma^{(2)}_x \cr &&& \cr
H2: & \sigma^{(2)}_y & \sigma^{(1)}_y &
\sigma^{(1)}_y \sigma^{(2)}_y \cr &&& \cr
H3: & \sigma^{(1)}_x \sigma^{(2)}_y & \sigma^{(2)}_x \sigma^{(1)}_y &
\sigma^{(1)}_z \sigma^{(2)}_z} \ .
\end{matrix}
\] 
where $\sigma^{(\alpha)}_x, \sigma^{(\alpha)}_y, \sigma^{(\alpha)}_z$
are the Pauli-matrices for the particle $\alpha, (\alpha = 1, 2)$. All
the three operators in either a row (denoted by H1, H2, H3) or a
column (denoted by V1, V2, V3) commute with each other. Further the
product of all the three operators in either a row or in column V1 and
V2 is equal to 1, while the product of the three operators in column
V3 equals (--1). Further the eigenvalues of all the nine operators are
+1 or --1. 

It follows that the product of the values $v (A)$ of all the
observalues in horizontal row and first and second vertical rows must
be equal to 1, and for the third vertical column it should be equal to
--1. This is however impossible since the product all the nine values of
operators, as calculated from the product of values of operators in
the horizontal rows comes out to 1, while the same product, as
calculated from the product of values of operators in the vertical
rows comes out to be -1. We have thus shown that such a noncontextual
value assignment to the physical observables in not possible, thus
proving Bell-Kochen-Specker theorem. This is a far simpler proof of
it, albeit for only system with a dimension of Hilbert space larger
equal to 4, while the original Bell-Kochen-Specker applied for systems
with Hilbert space dimension greater or equal to 3. 

We may mention here that, if we consider more generalised measurements
using ``Positive Operator Valued Measures'' (POVM's), instead of only
Von-Neumann measurements using only ``Projectors'', we can extend both
Gleasons and Kocher-Specker Theorems to Hilbert space of a qubits with
dimension 2. 

\subsection{Some Further Aspects}

\subsubsection{Stochastic Noncontextual Hidden Variable Theories}

As we have seen deterministic noncontextual hidden variable theories
are ruled out by Gleason-Bell and Kochen-Specker theorems. In quantum
theory however we have a kind of ``Statistical Noncontextuality'' in
that the expectation values of any observable is unchanged by a
simultaneous or previous measurement of observable which commute with
it. Roy and Singh therefore investigated whether a ``statistically
noncontextual'' hidden variable theory could reproduce this feature of
quantum mechanics. They showed that it is not possible to do so. 

\subsubsection{Finite Precision Measurements and Kochen-Specker
Theorem}

We saw that it was not possible to carry out
Kochen-Specker colouring of the surface of a unit phase. Note that we
are specifying each direction precisely with no latitude for
error. Meyer, in 1999, followed by Kent, raised the issue whether a
noncontexual hiddle variable model will allow us to reproduct the
predictions of quantum theory if we allow finite fixed precision in
experiments. Clifton and Kent later presented K-S colouring which
claim to do so over any dense subset. These colourings are extremely
discontinuous. There has been a lot of criticism of these claims and an
extended discussion of what they mean? Since any neighbourhood, no
matter how small, contains points coloured both red and green, it
follows that these models do not satisfy the ``faithful measurement
condition''. If an observable has a value, it is not guaranteed that
in a experiment, no matter how precise, we will obtain a value close
to it's value with a high probability. 

\section{Einstein-Locality and Quantum Mechanics}

\subsection{The Background}

The early phase of quantum theory beginning with Planck's work in 1900
on black-body radiation and ending with the discovery of quantum
mechanics by W. Heisenberg (1925), Paul Dirac (1925) and
E. Schr\"odinger (1926) is referred to as old quantum theory. In this
period Einstein beginning with his paper on ``light quantum
hypothesis'' in 1905, and Niels Bohr, beginning with his papers on
atomic structure and spectra in 1913, had been the main guiding lights
of quantum theory. After the discovery of the mathematical formation
of quantum mechanics in 1925-26, Niels Bohr, together with Heisenberg
and others, hammered out the ``Copenhagen Interpretation of quantum
mechanics'' which was to remain the dominant way to think about
quantum mechanics for a long time. Einstein, with his deep committment
to realism, did not subscribe to this interpretation. His initial
efforts were to see whether Heisenberg's uncertainty principle can be
circumvented and thereby showing the provisional nature of quantum
mechanics. These discussions have been summarised by Bohr in his
account of famous Einstein-Bohr dialogues. Einstein here conceded to
Bohr that this was not possible. Einstein however raised a much more
substantial argument, against the prevailing interpretation, which
depended on some unusual non-local aspects of two-particle systems in
quantum mechanics. 

\subsection{Einstein-Podolsky-Rosen Theorem}

In 1935, A. Einstein, B. Podolsky and N. Rosen (EPR) published a paper
``Can Quantum Mechanical Description of Reality be Considered
Complete?'' in Physical Review. It had a rather unusual title for a
paper for this journal. In view of this they provided the following
two definitions at the beginning of the paper: (1) A
\underbar{necessary} condition for the \underbar{completeness} of a
theory is that every element of the physical reality must have a
counterpart in the physical theory. (2) A \underbar{sufficient}
condition to identify an element of reality: ``If, without in any way
disturbing a system, we can predict with certainty (ie with
probability equal to unity) the value of a physical quantity, then
there exists an element of physical reality corresponding to this
physical quantity''. 

We now illustrate the use of these definitions for a single-particle
system. Let the position and momentum observable of the particle be
denoted by $Q$ and $P$ respectively. Since in an eigenstate of $Q$, we
can predict with certainty the value of $Q$, which is given by it's
eigenvalue in that eigenstate, it follows that the position $Q$ of the
particle is an element of physical reality (e.p.r.). Similarly the
momentum $P$ is also an e.p.r. The position $Q$ and the momentum $P$
however are not simultaneous e.p.r. So at the single particle level
there is no problem with quantum mechanics, as far as these
definitions of `completeness' and `elements of reality' are
concerned. 

The interesting new things are however encountered when a two particle
system is considered. Let the momenta and position of the two
particles be denoted respectively by $P_1$ and $Q_1$ for the first
particle and by $P_2$ and $Q_2$ for the second particle. Consider now
the two-particle system in the eigenstate of the relative-position
operator, $Q_2$--$Q_1$ with eigenvalue $q_0$. The relative position
$Q_2$--$Q_1$ can be predicted to have a value $q_0$ with probability
one in this state and thus qualifies to be an e.p.r. We can also
consider an eigenstate of the total momentum operator, $P_1 + P_2$,
with an eigenvalue $p_0$. The total momentum can be predicted to have
a value $p_0$ with probability one and thus also qualifies to be an
e.p.r. Furthermore relative position operator, $Q_2$--$Q_1$, and total
momentum operator, $P_1 + P_2$, commute with each other and thus can
have a common eigenstate, and thus qualify to be
\underbar{simultaneous} elements of physical reality. 

We consider the two-particle system in which two particles are flying
apart from each other having momenta in opposite directions and are
thus having a large spatial separation. The separation will be taken
so that no physical signal can reach between them. Let a measurement
of position be made on the first particle in the region $R_1$ and let
the result be $q_1$. It follows from standard quantum mechanics that
instantaneously the particle 2, which is a spatially for away region
$R_2$, would be in an eigenstate $q_0 + q_1$ of $Q_2$. The $Q_2$ is thus an
e.p.r. the position of second particle gets fixed to the value $q_0 +
q_1$ despite the fact that no signal can reach from region $R_1$ to
$R_2$ where the second particle is $A$ ``spooky action at a
distance'' indeed. On the other hand a
measurement of the momentum $P_1$ of the first particle, in the region
$R_1$ can be carried out and let it result in a measured value
$p_1$. It then follows from the standard quantum mechanics, that the
particle 2, in the region $R_2$ would be in an eigenstate of its
momentum $P_2$ with and eigenvalue $p_0$--$p_1$. The $P_2$ is thus
also an e.p.r. This however leads to a contradiction since $Q_2$ and
$P_2$ can not be a simultaneous e.p.r. as they do not commute. We
quote the resulting conclusion following from this argument as given
by Einstein in 1949,\\
\underbar{EPR Theorem}: The following two assertions are not
compatible with each other\\ 
``(1) the description by means of the $\psi$-function is complete \\
(2) the real states of spatially separated objects are independent of
each other''. 

The predilection of Einstein was that the second postulate, now
referred to as ``Einstein locality'' postulate, was true and thus EPR
theorem establishes the incompleteness of quantum mechanics. 

Einstein, Podolsky and Rosen were aware of a way out of the above
theorem but they rejected it as unreasonable. As they said ``Indeed
one would not arrive at our conclusion if one insisted that two or
more quantities can be regarded as simultaneous elements of reality
only when they can be simulateneously measured or predicted. On this
point of view, either one or the other, but not both simultaneously,
of the quantities $P$ and $Q$ can be predicted, they are not
simultaneously real. This makes the reality of $P$ and $Q$ depend upon
the process of measurement carried out on the first system, which does
not disturb the second system in any way. No reasonable definition of
reality could be expected to permit this''. 

\subsection{Einstein Locality}

We present here a selection of Einstein quotations on this important
concept. 

\begin{enumerate}
\item[{(i)}] \underbar{From his ``autobiographical notes'' (1949)}\\
``But on one supposition we should, in my opinion, absolutely hold
fast: the real factual situation of the system $S_2$ is independent of
what is done, with system $S_1$, which is spatially separated from the
formed''. 

\item[{(ii)}] \underbar{From Einstein-Born correspondence (March
1948)}\\
``That which really exists in B should $\cdots$ not depend on what
kind of measurement is carried out in part of space A; it should also
be independent of whether or not any measurement at all is carried out
in space A. If one adheres to this program, one can hardly consider
the quantum theoretical description as a complete representation of
the physically real. If one tries to do so in spite of this, one has
to assume that the physically real in B suffers a sudden change as a
result of a measurement in A. My instinct for physics bristles at
this''. 

\item[{(iii)}] \underbar{From Einstein-Born Correspondence (March
1947)} \\
``I can not seriously believe in (the quantumtheory) because it can
not be reconciled with the idea that physics should represent a
reality in time and space free from spooky actions at a distance
(Spukhafte Fernwirkungen)''. 
\end{enumerate}

\subsection{Bohm's Version of the EPR Analysis}

Einstein-Podolsky-Rosen discussion suffers from some inessential
mathematical technicalities, which were noted by a number of
authors. Some of these are, (i) the use of non normalisable plane-wave
eigenfunctions, (ii) neglect of time dependence of wave functions, and
(iii) non-self-adjointness of momentum operator over half-space. Bohm,
in order to present a clear account of the essential features of the
EPR analysis, reformulated it in terms of two spin-one-half
particles. This formalism has also been influencial in later work of
John Bell and others on the foundations of quantum mechanics. We
therefore present it now. 

Consider a spin one-half particle with spin angular momentum $\vec S =
{1\over 2} \vec\sigma \hbar$. Here $\vec\sigma$ are the usual Pauli
matrices. From now we will often set $\hbar = 1$. The component of
spin along the diretion unit-vector $\hat n$ is given by $\vec S \cdot
\hat n$. Let us denote it's eigenvectors by $\chi_m (\hat n)$ with $m$
specifying the eigenvalue ie
\[
\vec S \cdot \hat n \ \ \chi_m (\hat n) = m \ \ \chi_m (\hat n) .
\]
Here $m$ can take only two values given by $m = {1\over 2}$,
corresponding to spin pointing up in $\hat n$ direction and $m =
-{1\over 2}$ corresponding to spin pointing down in $\hat n$
direction. We also denote $S \cdot \hat z$ by $S_z$ and $S \cdot \hat
x$ by $S_x$. In an eigenstate of $S_z$, we can predict with certainty
it's value and thus it is an e.p.r. Similarly one can see that $S_x$
is also an e.p.r. They are however not simultaneous e.p.r. as they do
not commuts. 

We now consider, a two spin one half particle system in a singlet
state ie which has total spin-angular-momentum, $S^{(1)} + S^{(2)}$,
equal to zero. We will use superscripts 1 and 2 to denote the
quantities referring the particle 1 and 2 respectively. We will also
ignore the inessential space degree of freedom. The eigenfunction of
the singlet state is given by 
\[
\psi ({\rm singlet}) = {1\over \sqrt{2}} \left[\chi^{(1)}_{1/2} (\hat
z) \chi^{(2)}_{-1/2} (\hat z) - \chi^{(1)}_{-1/2} (\hat z)
\chi^{(2)}_{1/2} (\hat z) \right] .
\]

We now imagine this singlet state two-particle state, at rest, decays
into it's two constituents, which fly apart with opposite momenta. Let
the particle 1 and 2 be respectively in regions $R_1$ and $R_2$ which
are far apart. If we now measure $S^{(1)}_z$ for particle 1 in the
region $R_1$ we will obtain a definite value which is $+{1\over 2}$ or
$-{1\over 2}$. We can therefore assert with certainty that $S^{(2)}_z$
has a definite value, which is negative of the value found for
$S^{(1)}_z$, in the far away region $R_2$. 

We could however have decided to measure $S^{(1)}_x$ instead of
$S^{(1)}_z$. The singlet eigenfunction can be also rexpressed as 
\[
\psi ({\rm singlet}) = {1\over \sqrt{2}} \left[\chi^{(1)}_{1/2} (\hat
x) \chi^{(2)}_{-1/2} (\hat x) - \chi^{(1)}_{-1/2} (\hat x)
\chi^{(2)}_{1/2} (\hat x) \right] .
\]
A measurement of $S^{(1)}_x$ in the region $R_1$ would lead to a
definite value for $S^{(1)}_x$ which is either $+{1\over 2}$ or
$-{1\over 2}$. We can therefore assert with certainty that the value
of $S^{(2)}_x$ also has a definite value, which is negative of
$S^{(1)}_x$ value found in $R_1$, in the far away region
$R_2$. However the particle 2 can not have a definite value for both
$S^{(2)}_z$ and $S^{(2)}_x$ since these do not commute. The EPR
theorem, therefore, follows. Surprising element here is not that the
value of $S^{(2)}_z$ (or $S^{(1)}_x$) is negative to that of
$S^{(1)}_z$ (or $S^{(1)}_x$), but rather that the far away particle 2
is in an eigenstate of $S^{(2)}_z$ or $S^{(2)}_x$ depending on what we
decide to measure in region $R_1$. As Greenberger and Yasin said
``Reality should be made of sterner stuff''. 

\subsection{Bell's Inequalities and Bell's Theorem}

As a consequence of E.P.R. work we see Einstein nonlocality is a
feature of quantum mechanics. On the other hand Einstein-locality
looks a very nice and desirable feature to have in a basic theory of
nature. Can we retain Einstein-locality if we admit of a hidden
variable substratum to quantum mechanics? John Bell investigated that
possibility by giving concept of Einstein-locality a precise
mathematics formulation. 

Let the set of hidden variables be denoted by $\lambda$. We take then
to be distributed probability distribution $\rho (\lambda)$ which is
positive semidefinite and is normalised to unity. We use Bohm
formulation of EPR work and we shall be measuring correlations of spin
components $\sigma^{(1)} \cdot \hat a$ and $\sigma^{(2)} \cdot \hat b$
for the two spin ${1\over 2}$ particles. Let the values fixed by
hidden variables $\lambda$ for $\sigma^{(1)} \cdot \hat a$ be denoted
by $A (\hat a, \lambda; b)$ while for $\sigma^{(2)} \cdot \hat b$ be
denoted by $B (\hat b, \lambda; a)$. since the observable values of
$\sigma^{(1)} \cdot \hat a$ and $\sigma^{(2)} \cdot \hat b$ are their
eigenvalues, which are $+1$ or $-1$, we have to take allowed value of
$A (\hat a, \lambda; b)$ and $B(\hat b, \lambda; a)$ to also be $+1$
or $-1$. All the quantities $\rho (\lambda), A (\hat a, \lambda; \hat
b)$ and $B (\hat b, \lambda; \hat a)$ will have dependenace on the
wavefunction $\psi$ of the system as well which we donot explicitly
indicate. In the hidden variable theory $P (\hat a, \hat b)$, ie the
expectation value of the product of $\sigma^{(1)} \cdot \hat a$ and
$\sigma^{(2)} \cdot \hat b$, would be given by
\[
P_{h.v} (\hat a, \hat b) = \int d\lambda \rho (\lambda) \ A (\hat a,
\lambda; \hat b) B(\hat b, \lambda; \hat a) .
\]

We have so far not used the concept of Einstein locality. In it's
spirit we assume, following Bell, that the hidden variable value
$A(\hat a, \lambda; \hat b)$ of the observable $\sigma^{(1)} \cdot
\hat a$ should not depend on the setting $\hat b$ of the Stern-Gerlach
magnet used to measure $\sigma^{(2)} \cdot \hat b$ in the far away
region ie
\[
A (\hat a, \lambda; \hat b) = A (\hat a, \lambda) .
\]
Similarly
\[
B (\hat b, \lambda; \hat a) = B (\hat b, \lambda). 
\]
So, in an Einstein-locality obeying hidden variable theory
\[
P_{h.v} (\hat a, \hat b) = \int d \lambda \rho (\lambda) A (\hat a,
\lambda) B(\hat b, \lambda) 
\]
where
\begin{eqnarray*}
|A (\hat a, \lambda) | &=& 1 \\
|B (\hat b, \lambda) | &=& 1 \\
{\rm and} \ \ \rho(\lambda) \geq 0, & & \int d \lambda \rho
(\lambda) = 1 .
\end{eqnarray*}

The correlation coefficient $P (\hat a, \hat b)$ is given in quantum
mechanics by 
\begin{eqnarray*}
P_{QM} (\hat a, \hat b) &=& \langle \psi | \sigma^{(1)} \cdot \hat a \
\sigma^{(2)} \cdot \hat b | \psi \rangle \\
&=& - \hat a \cdot \hat b \ \ \ \ {\rm for} \ \psi = \psi ({\rm
singlet}) .
\end{eqnarray*}
Note that in quantum mechanics we have perfect anticorrelation given
by
\[
P_{QM} (\hat a, \hat a) = -1 \ \ \ \ {\rm for} \ \psi = \psi ({\rm
singlet}) .
\]
If we specialise to the case of singlet wavefunction and demand
perfect anticorrelation, an extra assumption, for the correlation
function for hidden variable theories it
\[
P_{h.v} (\hat a, \hat a) = -1 
\]
we must have 
\[
B (\hat a, \lambda) = - A (\hat a, \lambda) .
\]
We then have the symmetric expression 
\[
P_{h.v} (\hat a,\hat b) = - \int d \lambda \rho (\lambda) A (\hat a,
\lambda) A (\hat b, \lambda) .
\]
It is easy to see that
\[
|P_{h.v} (\hat a, \hat b) - P_{h.v} (\hat a, \hat c) | \leq 1 +
P_{h.v.} (\hat b, \hat c) 
\]
\underbar{\bf Proof}: 
\begin{eqnarray*}
P_{h.v} (\hat a, \hat b) - P_{h.v} (\hat a, \hat c) &=& - \int d
\lambda \rho (\lambda) A (\hat a, \lambda) \left[A(\hat b, \lambda) -
A (\hat c, \lambda) \right] , \\
|P_{h.v} (\hat a, \hat b) - P_{h.v} (\hat a, \hat c) | &\leq& \int d
\lambda \rho(\lambda) | A(\hat a, \lambda) A(\hat b, \lambda) | \left[
1 - A(\hat b, \lambda) A(\hat c, \lambda)\right] \\
&\leq& \int d \lambda \rho (\lambda) \left[ 1 - A(\hat b, \lambda) A
(\hat c, \lambda) \right] = 1 + P_{h.v} (\hat b, \hat c). 
\end{eqnarray*}
We can, infact, derive four inequalities in a similar way,
\[
1 \geq \eta_a \eta_b P_{h.v} (\hat a, \hat b) + \eta_a \eta_c P_{h.v}
(\hat a, \hat c) + \eta_b \eta_c P_{h.v} (\hat b, \hat c)
\]
where $(\eta_a)^2 = (\eta_b)^2 = (\eta_c)^2 = 1$. These were the first
inequalities, following from Einstein-locality for the correlation
coefficients, which were derived by John Bell. All such inequalities
on correlation coefficients are now known, generically, as Bell's
inequalities. 

Bell made the remarkable discovery that these inequalities are not
always consistent with the prediction of quantum-mechanics, $P_{Q.M}
(\hat a, \hat b) = - \hat a \cdot \hat b$ given above. For example let
$\hat a = (1, 0, 0), \hat b = (-{1\over 2}, {\sqrt{3}\over 2}, 0),
\hat c = (-{1\over 2}, -{\sqrt{3}\over 2}, 0)$ and choose $\eta_a =
\eta_b = \eta_c = 1$. We then have $\hat a \cdot \hat b = \hat a \cdot
\hat c = \hat b \cdot \hat c = - {1\over 2}$ and thus Bell's
inequality, given above becomes 
\[
1 \geq {1\over 2} + {1\over 2} + {1\over 2} = {3\over 2} 
\]
which is clearly not satisfied. It follows that,\\
\underbar{Bell's Theorem}:\\
Quantum mechanics is inconsistent with a Einstein-locality
obeying hidden variable theories. \\
Stapp has called it the most profound discovery of twentieth century
physics. 

The above discussion can be generalised to include additional hidden
variables for the measuring instruments as well provided the
distribution of these hidden variables does not depend on the setting
of the far away measuring instrument. Denoting by $\bar A (\hat a,
\lambda)$ and $\bar B (\hat b, \lambda)$, the averaging of $A (\hat a,
\lambda)$ and $B(\hat b, \lambda)$ respectively, we have 
\[
|\bar A (\hat a, \lambda) | \leq 1, \ \ \ |\bar B (\hat b, \lambda)|
\leq 1 
\]
and we have to use these averaged quantities $\bar A (\hat a,
\lambda)$ and $\bar B (\hat b, \lambda)$, instead of $A (\hat a,
\lambda)$ and $B (\hat b, \lambda)$ respectively in the expression for
$P_{h.v} (\hat a, \hat b)$ given earlier. The same Bell's inequalities
still follow. 

\subsection{Clauser-Horne-Shimony-Holt (CHSH) Form of Bell's
Inequalities} 

The assumption of perfect correlation $P(\hat a, \hat a) = -1$ for a
singlet two spin one-half particles may be hard to meet in practice as
it would require ideal (and may be identical) detectors on both
sides. The CHSH formulation does not assume this. Using
\[
P_{h.v} (\hat a, \hat b) = \int d \lambda \rho (\lambda) \bar A (\hat
a, \lambda) \bar B (\hat b, \lambda)
\]
with
\begin{eqnarray*}
| \bar A (\hat a, \lambda) | \leq 1, && | \bar B (\hat b, \lambda) |
\leq 1, \\
\rho (\lambda) \geq 0, && \int d \lambda \rho (\lambda) = 1 ,
\end{eqnarray*}
it is possible to derive the following Bell-CHSH equalities
\[
S = | P_{h.v} (\hat a, \hat b) - P_{h.v} (\hat a, \hat b') | +
|P_{h.v} (\hat a', \hat b) + P_{h.v} (\hat a', \hat b') | \leq 2 . 
\]
In fact we have the same inequalities valid even if 
\[
P(\hat a_i, \hat b_j) \to Q (\hat a_i, \hat b_j) = \eta_i \eta_j P
(\hat a_i, \hat b_j) 
\]
where
\[
(\hat a_1, \hat a_2, \hat b_1, \hat b_2) = (\hat a, \hat a', \hat b,
\hat b') \ {\rm} \ \eta^2_i = \eta^2_j = 1 .
\]
The proof is similar to Bell's inequalities given earlier. 

If these inequalities are consistent with Quantum-mechanics we should
have 
\[
|\hat a \cdot \hat b - \hat a \cdot \hat b'| + |\hat a' \cdot \hat b +
\hat a' \cdot \hat b'| \leq 2
\]
for all possible choices of unit vectors. Choosing 
\[
\hat a = (0, 1, 0), \hat b = ({1\over \sqrt{2}}, {1\over \sqrt{2}},
0), \hat a' = (1, 0, 0), \hat b' = ({1\over \sqrt{2}}, -{1\over
\sqrt{2}}, 0)
\]
we see that lefthand side is equal to $2\sqrt{2}$ while the right hand
side of the inequality is equal to 2. We thus see that quantum
mechanics is in general not consistent with hidden variable
Einstein-local theories. The violation, which is maximum possible, is
a factor of $\sqrt{2}$ here. 

\subsection{Wigner's Proof of Bell-CHSH Inequalities}

We should also refer to a proof of Bell's inequalities by Wigner. His
assumption are (i) physical realism: spin components of the two  
particles have definite preassigned values even for non-commuting
observables, and (ii) Einstein-locality: a measurement on a spin
component on one side does not modify the preassigned value of the far
away spin component. In Wigner's proof probabilities appear for the
first time. Let $w (s, s', t, t')$ be the fraction of population of
those particles for which preassigned values are given by
\[
A(\hat a) = s, \ A(\hat a') = s', \ B(\hat b) = t \ {\rm and} \ B
(\hat b') = t' .
\]
Here $s, s', t, t'$ are all equal to $\pm$ 1. We then have
\[
\sum w (s, s', t, t') = 1, \ \ \ w (s, s', t, t') \geq 0 , 
\]
\begin{eqnarray*}
P (\hat a, \hat b) &=& \sum w (s, s', t, t') st , \\
P (\hat a, b') &=& \sum w (s, s', t, t') st' , \\
P (\hat a', \hat b) &=& \sum w (s, s', t, t') s't , \\
P (\hat a', \hat b') &=& \sum w (s, s', t, t') s't' . 
\end{eqnarray*}
Again it is easy to derive Bell-CHSH inequality using these
expression. Note that probability concept is used here in the same way
as in classical statistical physics. 

\subsection{Experimental Tests of Bell's Inequalities}

The Bell-CHSH inequalities, published in 1969, were amenable to
experimental tests and showed the possibility of experimental tests
for discriminating between hidden variable theories obeying Einstein
locality and quantum mechanics. The first generation experiments were
carried in nineteen seventies. Better second generations experiments
were carried out by Alain Asp\'ect at Orsay in 1980-82. For the
settings, where the conflict is maximu, it is found that 
\[
S_{\rm exp} = 2.697 \pm 0.015
\]
The quantum mechanical prediction is
\[
S_{\rm QM} = 2.70 \pm 0.05 .
\]
The uncertainties in $S_{\rm QM}$ refer to slight lack of symmetry,
about $\pm 1\%$, of two channels of polarisers used in the experiment
(ideally $S_{\rm QM} = 2\sqrt{2}$). The local hidden variable theory
has, as follows for CHSH inequality
\[
| S_{h.v} | \leq 2.0 
\]
It would seem that experiments are in favour of quantum mechanics and
Einstein nonlocality is a feature of nature. 

More precise third generation experiments started in late ninetten
eightees and still are in progress. They would take care of various
loopholes in earlier experiments such as (i) low detection
efficiencies of photon detectors, (ii) possibility of light signalling
between the detectors by having polarisers which can be randomly oriented
in timing faster than the time taken by light to reach from one
polariser to another. 

\subsection{Bell's Theorem without Inequalities}

\subsubsection{Greenberger-Horne-Zeilinger (GHZ) Proof}

It was realised by Greenberger, Horne and Zeilinger that for a three
spin-${1\over 2}$ particle system, it is possible to demonstrate a
conflict between local realism and quantum mechanics more directly
without using Bell's inequalities. 

Consider the three spin-${1\over 2}$ particle system in the state
\[
\psi = {1\over \sqrt{2}} [ | \uparrow\uparrow\uparrow \rangle - |
\downarrow \downarrow \downarrow \rangle ] 
\]
where $|\uparrow\uparrow\uparrow \rangle (|\downarrow \downarrow
\downarrow)$ denotes the state with all the three particles have $z$
component of the spin $S_z = {1\over 2} \sigma_z$ equal to $+{1\over
2} (-{1\over 2})$.  It is easy to verify that
\begin{eqnarray*}
\sigma^{(1)}_x \sigma^{(2)}_y \sigma^{(3)}_y | \psi \rangle = | \psi
\rangle \\  
\sigma^{(1)}_y \sigma^{(2)}_x \sigma^{(3)}_y | \psi \rangle = | \psi
\rangle \\  
\sigma^{(1)}_y \sigma^{(2)}_y \sigma^{(3)}_x | \psi \rangle = | \psi
\rangle . 
\end{eqnarray*}  
Here superscipts will refer to the particles. Let the three particle
separate and be in a far away regions. We can measure $\sigma^{(2)}_y$
with a result $m^{(2)}_y$, and $\sigma^{(3)}_y$ with a result
$m^{(3)}_y$, we can predict the value $m^{(1)}_x$ of $\sigma^{(1)}_x$
with certainty. It would be given by
\[
m^{(1)}_x m^{(2)}_y m^{(3)}_y = 1. 
\]
Thus $\sigma^{(1)}_x$ would be an e.p.r. Similarly the value
$m^{(1)}_y$ of $\sigma^{(1)}_y$ can be predicted with certainty by
measuring the value $m^{(2)}_x$ of $\sigma^{(2)}_x$ and the value
$m^{(3)}_y$ of $\sigma^{(3)}_y$ in the far away regions. It would be
given by
\[
m^{(1)}_y m^{(2)}_x m^{(3)}_y = 1. 
\]
Thus $\sigma^{(1)}_y$ is also an e.p.r. Similarly $\sigma^{(2)}_x,
\sigma^{(2)}_y, \sigma^{(3)}_x$ and $\sigma^{(3)}y$ are also e.p.r. We
have one more relation
\[
m^{(1)}_y m^{(2)}_y m^{(3)}_x = 1.
\]
By multiplying these three relations it follows that
\[
m^{(1)}_x m^{(2)}_x m^{(3)}_x = 1. 
\]
Since $(m^{(i)}_y)^2 = 1$ for $i = 1,2,3$. But this is not consistent
with quantum mechanics, since we also have
\[
\sigma^{(1)}_x \sigma^{(2)}_x \sigma^{(3)}_x |\psi \rangle = (-1)
|\psi \rangle 
\]
which would imply in a local realist theory
\[
m^{(1)}_x m^{(2)}_x m^{(3)}_x = -1.
\]
This argument demonstrate the conflict between deterministic
local-hidden variable theories. Of course Bell's argument is
generalisable to some forms of Stochastic local-hidden variable
theories as well. 

\subsubsection{Hardy's Version of EPR Correlations}

L. Hardy found that, surprisingly there exist two spin one-half
particle states for which it is possible to demonstrate the conflict
between local realism and quantums mechanics. We first present it in
a paraphase by Stapp. 

Consider again two spin one half particles moving away from each other
towards region $A$ and $B$ which are far from each other. 

We denote the two physical observable, to be measured in each region
$A$ and $B$ by ``size'' and ``colour''. The ``size'' can take two
values viz ``large'' and ``small''. The ``colour'' can be ``black'' or
``while''. The constructed state has the following perfect corelations
built in
\begin{enumerate}
\item[{(i)}] If ``size'' was measured in region $A$ and was found to
have the value ``large'', then if ``colour'' was measured in region
$B$ it would be found to be ``white'' with probability equal to 1 ie

\vspace{-.1cm} 
[region $A$: ``size'' = ``large''] $\to$ [region $B$: ``colour'' =
``white''], 
\item[{(ii)}] [region $B$: ``colour'' = ``white''] $\to$ [region $A$:
``colour'' = ``black''], 
\item[{(iii)}] [region $A$: ``colour'' = ``black''] $\to$ [region $B$:
``size'' = ``small'']\\
Would it now be correct to conclude that the following correlation
should also be perfect, ie probability equal to 1,

\vspace{-.1cm}
[region $A$: ``size'' = ``large''] $\to$ [region $B$: ``size'' =
``small''], \\
as it seems to be implied by classical physics and common sense. Hardy
showed that, in a quantum world, this is not necessarly so but one can
find, with a probability of upto about 9\%, the correlation

\vspace{-.1cm}
[region $A$: ``size'' = ``large''] $\to$ [region $B$: ``size'' =
``large'']. 
\end{enumerate}

We now present briefly some mathematical details. Consider the
observables $U^{(1)}$ (and $U^{(2)}$) for the particle 1 (and 2) given
by
\[
U^{(i)} = |u^{(i)} \rangle \langle u^{(i)} |
\]
where $|u^{(i)}\rangle$ and $|v^{(i)}\rangle$ forms an orthonormal
basis states for particle $i(=1,2)$. Consider also another observables
$U^{(i)\prime} (i=1,2)$ for the two particles given by
\[
U^{(i)\prime} = |u^{(i)\prime} \rangle \langle u^{(i)\prime} | 
\]
where $|u^{(i)\prime}\rangle$ and $|v^{(i)\prime}\rangle$ forms
another orthonormal basis states of the particle $i(=1,2)$. All these
observables are projection operators with eigenvalues eqal to 0 or 1. 

Hardy discovered that two spin-${1\over 2}$ particles states
$|\psi\rangle$ existswhich satisfy, (with $V^{(i)\prime} = 1 -
U^{(i)}$) 
\begin{enumerate}
\item[{(i)}] $U^{(1)} U^{(2)} |\psi\rangle = 0$ 
\item[{(ii)}] $(1-U^{(1)}) V^{(2)\prime} |\psi\rangle = 0$ \ ie
$U^{(2)\prime} = 0 \Longrightarrow U^{(1)} = 1$ 
\item[{(iii)}] $V^{(1)\prime} (1-U^{(2)}) |\psi\rangle = 0$ \ ie
$U^{(2)} = 0 \Longrightarrow U^{(1)\prime} = 1$ 
\item[{(iv)}] there is a non-zero probability $p$ for finding
$U^{(1)\prime} = U^{(2)\prime} = 0$. \\
More explicitly, we have 
{\small
\begin{eqnarray*}
\sqrt{1 - p_1p_2} |\psi\rangle &=& \sqrt{(1-p_1)(1-p_2)} |v^1,v^2
\rangle - \sqrt{p_1(1-p_2)} |u^1,v^2 \rangle - \sqrt{p_2(1-p_1)} |
v^1, u^2 \rangle \\[2mm]
p &=& {p_1 (1 - p_1) (p_2) (1 - p_2) \over 1 - p_1p_2}
\end{eqnarray*}}
where $p_1, p_2$ are real numbers lying between 0 and 1.  
\end{enumerate} 

The maximum value of $p$ is obtained by choosing $p_1 = p_2 = {1\over
\tau}$ where $\tau = {\sqrt{5} + 1 \over 2}$ is the golden ratio. We
thus have 
\[
0 \leq p \leq {1\over \tau^5} \approx 0.09017 \cdots 
\]

By using Hardy ladders, ie by a consideration of $N$ observables, each
two valued, in each region $A$ and $B$, the probability of violation
of local realism $p_N$ can be made as large as ${1\over 2}$ by letting
$N \to \infty$. 

\subsection{Superluminal Signalling} 

Quantum mechanics satisfies a signal locality, ie no faster than light
signalling, requirement for statistical averages. Let $B$ denote a
physical observales being measured in a space like region separated
from the region to which a varibale $A$ refers. On measuring $B$ the
density matrix $\rho$ changes to $\rho'$, ie 
\[
\rho \to \rho' = \sum_\beta P_\beta \rho P_\beta 
\]
where $P_\beta$'s are projectors to different eigenstates of the
observable $B$. So if $A$ is measured after $B$ is measured first, we
get the expection value $\langle A \rangle$ of $A$, 
\begin{eqnarray*}
\langle A \rangle &=& {\rm Tr} (\rho' A) = {\rm Tr} (\sum_\beta
P_\beta \rho P_\beta A) \\
&=& {\rm Tr} \sum_\beta (P_\beta \rho A P_\beta)^- 
\end{eqnarray*}
since $[A, P_\beta] = 0$ as $[A, B] = 0$. Using cyclicity of trace
and general properties of projectors 
\[
P^2_\beta = P_\beta , \ \ \sum_\beta P_\beta = 1 ,
\]
we finally obtain 
\[
\langle A \rangle = {\rm Tr} (\rho A) .
\]
This however is also the expectation value of $A$ even if $B$ had not
been measured earlier. We can not use the earlier measurement of $B$
to do any signalling as long as only expectation values are measured. 

It was shown that requirement of signal locality can also be
formulated for local hidden variable theories and in general leads to
testable inequalities. 

It might appear as if EPR correlations might lead to superliminal
signalling. This is not so, since the sequence of measurements of a
spin component in a region produces a random sequence, which not
having any structure, can not contain information. 

\subsection{More Bell's Inequalities}

The context of Einstein locality is not exhausted, even for two
spin-${1\over 2}$ particle system, by the Bell's inequalities given by
Bell himself or by CHSH discussed earlier. A large number of these
have written down for two spin-${1\over 2}$ particle system, multiple
particle system and higher spin systems. They have also been written
for phase space.

\section{Envoi}

The subject of ``foundations of quantum mechanics'', which includes
EPR correlation, was somewhat philosophical and generally hard-boiled
physicist would turn their noses at it. John Bell's work in early
sixties showed that these philosophical discussion can be subject to
precise experimentation. If the experiments had been found not to
violate Bell's inequalities, quantum mechanics would have been in
serious trouble. The Einstein nonlocal nature of quantum mechanics,
which was exposed through EPR correlations, have been found in last
two decades to be, far from being an embarrasement, a resource in many
technical engineering applications. These applications include newly
emerging areas of quantum cryptography and quantum teleportation. The
whole area of quantum information and computing is intensely
active. One has come a long way from philosophy to technology.

\newpage 

\section{Bibiliographical Notes}

\end{document}